# A Simple Software Application for Simulating Commercially Available Solar Panels

[1*] Nalika Ulapane, [2] Sunil Abeyratne, [3] Prabath Binduhewa, [4] Chamari Dhanapala, [5] Shyama Wickramasinghe, [6] Nimal Rathnayake

[1] Centre of Excellence for Autonomous Systems, University of Technology Sydney,

[1] A.M.Ulapane@student.uts.edu.au

[2, 3, 4, 5, 6] Department of Electrical & Electronic Engineering, University of Peradeniya,

[2] sunil@ee.pdn.ac.lk, [3] prabathb@ee.pdn.ac.lk, [4] chamari4d@gmail.com,

[5] shyama.w06@gmail.com, [6] nimal@ee.pdn.ac.lk

*Abstract.* **This article addresses the formulation and validation of a simple PC based software application developed for simulating commercially available solar panels. The important feature of this application is its capability to produce speedy results in the form of solar panel output characteristics at given environmental conditions by using minimal input data. Besides, it is able to deliver critical information about the maximum power point of the panel at a given environmental condition in quick succession. The application is based on a standard equation which governs solar panels and works by means of estimating unknown parameters in the equation to fit a given solar panel. The process of parameter estimation is described in detail with the aid of equations and data of a commercial solar panel. A validation of obtained results for commercial solar panels is also presented by comparing the panel manufacturers' results with the results generated by the application. In addition, implications of the obtained results are discussed along with possible improvements to the developed software application.**

## 1. Introduction

Solar panels are increasingly used renewable energy sources which are widely utilized from small scale domestic applications to powering up large scale smart grids [1]. Therefore, a large variety of solar panels with various voltage and power ratings are available at commercial level[2 - 5].

A concern associated with solar power is that its costs still being relatively higher than conventional energy technologies even though its costs have reduced rapidly in the recent past [6]. Apart from that, the high variation shown by solar panel output voltages, currents and powers due to environmental conditions such as temperature and solar irradiance is also classified as a disadvantage [1, 7]. Due to these reasons, it is extremely important to get an idea whether a solar power plant will be the optimal solution for the energy requirement of a certain location, before actually investing in constructing a plant. If such pre - analysis is not conducted, there is always a risk in investments in solar power plants resulting in losses as the pay back periods may end up being longer than the actual plant lifetimes.

To facilitate the pre - analysis of solar power plants, many researchers have formulated methods in simulating and emulating solar panels and solar cells. These methods include software simulation techniques which model solar cells using different mathematical models [8 - 13]. However, the focus of these researches has been in developing accurate mathematical models to simulate solar cells and not targeted in simulating commercially available solar panels. On the other hand, some efforts have also been made in constructing power supplies which behave similar to solar cells using innovative power electronic topologies [14, 15]. But, these designs too are not for the purpose of emulating commercial solar panels. The best approaches in addressing this issue have been developed as commercial products known as solar emulators [16]. These happen to be programmable power supplies which behave similar to commercial solar panels when required user inputs are provided. However, these power supplies are also quite expensive to date and thus, provide room for more improvement. Due to the above reasons, the requirement for a simple, cost effective and accurate simulation application which can handle commercially available solar panels is unanimously risen and this was the motivation behind this research.

The objective of this research was to design a simple and effective solar panel simulation software application and this article addresses the approach followed. The proposed application overcomes the limitations of the software simulation techniques in literature by enabling the capability of simulating not only solar cells, but also commercially available solar panels. The basis of this application is the standard governing equation for solar panels which relates the output current of a solar panel ($I$) to the output panel voltage ($V$) as expressed in equation (1) [1,7].

$$I = I_{sc} - I_0 \left\{ \exp\left[ q\left( \frac{V + IR_s}{nkT} \right) \frac{1}{N} \right] - 1 \right\} \qquad (1)$$

Equation (1) is derived from the standard single diode solar cell model and the symbols $I_{sc}$, $I_0$, $R_s$, $N$, $T$, $n$, $k$ and $q$ are the short circuit current, diode reverse saturation current, series resistance, number of solar cells in the panel, cell temperature in Kelvin, diode ideality factor, Boltzmann's constant (1.381E-23 J/K) and charge of an electron (1.602E-19 C) respectively [1, 7]. As only the equation is of importance for the context of this article, the derivation of it from the mathematical model is not addressed here. But, interested scholars can get a detailed description of its derivative by referring [1].

The developed application takes seven typical values specified in a datasheet of a solar panel that needs to be simulated, along with the desired environmental conditions (solar irradiance and cell temperature) as user inputs. Those happen to be the only inputs required to get an output with appreciable accuracy from the novel application. Thus, a minimal input data requirement exists for this method and this is a great advantage in contrast to some existing data demanding solar panel simulation techniques discussed in [1] and [7 - 13]. After acquiring the required data, the application performs a quick estimation of some unknowns in equation (1). When the unknowns are estimated, the application uses equation (1) to generate two plots showing the variation of panel output current and power against panel output voltage. Soon after that, the application tracks the maximum power that can be delivered at the given environmental condition, it displays the maximum power along with the corresponding voltage and current to the user. In technical terms, this means that the application is capable of performing a maximum power point tracking (mppt) as well. This is very useful in the context of optimal usage of solar power as discussed in [17 - 19].

The above paragraph basically summarizes the operation of the developed software application and a detailed discussion regarding this is carried out in the sections to follow. The underlying algorithm of this application is by no means optimized or developed to commercial standards. However, it is efficient enough to simultaneously produce full solar panel output characteristic curves, full power profiles and the maximum power point information in a matter of a few fractions of a second when run using MATLAB with moderate processing power. This is an indication of the potential possessed by this application to be useful in real time solar panel monitoring and controlling tasks. The application as it is now, can be classified as a flexible research tool which enables researchers to simulate the output characteristics of commercial solar panels or solar cells by simply feeding few initial data and specifying any desired environmental condition. Thus, a main advantage in the proposed application is its minimal data requirement and this aspect greatly increases its user friendliness.

This article from here onwards is organized in such a way that it discusses about the data required for the process followed by an overview of the software user interface and the underlying algorithm. Since the algorithm is based on a parameter estimation, a detailed description of the estimation process along with derivation of mathematical equations is included next. Then, the method of generating solar panel output curves by taking environmental conditions into consideration is described in the following section. These solar panel output curves happen to be the useful result generated by the application. This section is followed by a validation of the results generated by the application. The validation is performed by means of comparing simulation results of two commercially available solar panels with the data provided by their manufacturers. The selected solar panels for the purpose were the BP SX 150 [2] and BP 3210 N [3] models. Finally, a comprehensive implication of the results is discussed while highlighting the room for improvement in the overall process.

## 2. Required input data for simulating a commercially available solar panel

As mentioned in the introduction, the minimal input data requirement is an advantage in the proposed method. The required data can be extracted from any standard solar panel data sheet. These data values are usually specified by solar panel manufacturers at standard test conditions (STC) which are considered as 1000 W/m$^2$ solar irradiance and 25 $^o$C cell temperature [1, 2, 7]. The BP SX 150 solar panel was selected for demonstrating the simulation in this article. The reason for selecting this panel was the fact that the panel manufacturer providing several solar panel output characteristic curves at different cell temperatures in the solar panel data sheet [2]. This enabled in validating this software application through comparing the manufacturer's output characteristics with the simulation results.

An example for a collection of the seven data values required for this simulation process is provided in Table 1. The data in Table 1 were extracted from the BP SX 150 solar panel data sheet [2]. It should be noted that two distinct categories can be identified among these data values. The first four data values belong to one category and they are specified at STC. This means that these happen to be parameters and they will have different values when the environmental conditions are different from STC. But, the manufacturers usually specify the values of theses parameters at STC, in the respective solar panel data sheets. The last three data values belong to the second category and they happen to be constants for the solar panel irrespective of environmental conditions. Symbols used for these values in derivations in the sections to come, are provided in the second column.

Table 1. Manufacturer specified values of the BP SX 150 solar panel

| Category | Symbol | Parameter | Value |
|---|---|---|---|
| Category 1 | $V_{oc}\|_{stc}$ | Open - circuit voltage at STC | 43·5 V |
| | $I_{sc}\|_{stc}$ | Short - circuit current at STC | 4·75 A |
| | $V_{mp}\|_{stc}$ | Maximum power point voltage at STC | 34·5 V |
| | $I_{mp}\|_{stc}$ | Maximum power point current at STC | 4·35 A |
| Category 2 | $N$ | Number of solar cells in the panel | 72 |
| | $a$ | Fractional temperature coefficient of short circuit current | 0·00065 /°C |
| | $b$ | Fractional temperature coefficient of open circuit voltage | -0·16 V/°C |

Once a data set similar to that in Table 1 along with environmental conditions are fed in, the developed application sequentially functions to estimate unknowns in equation (1), generate plots showing panel output characteristics, track maximum power point for the given environmental condition and finally display everything to the user.

## 3. The developed software interface for simulating solar panels

This section covers details about the software interface developed for the purpose and it is shown in Figure 1. A similar interface can be developed using any suitable software, but the one shown in Figure 1 was generated using the MATLAB GUI (Graphical User Interface) tool. The way to use the interface is described in brief from here onwards.

When the associated code is run, the whole interface appears with blank spaces and titles and the process should be initiated from the "Panel of Parameters". Seven blank spaces appear on the panel for the user to enter the values taken from the desired solar panel data sheet, similar to those in Table 1. In the interface in Figure 1, the values given in Table 1 are actually entered. Soon after entering the data, the "Estimate Parameters" tab should be pressed. When moderate processing power is used, within a few fractions of a second after pressing the tab, the values estimated by the algorithm for the unknowns in equation (1) are displayed to the user in the "Estimated Parameters Per Cell" sections. More about the unknown parameters in equation (1) and the parameter estimation process is discussed in Sections 4 and 5. Reviewers might argue about the significance of the mediocre statements "moderate processing power" and "a few fractions of a second". But, it should be mentioned that the statement about the time taken to estimate and display the results (processing time) is made purely by visual inspection. As said in the introduction, the underlying algorithm of this application is by no means optimized and highlighting the optimized efficiency of the algorithm is not the purpose of this article. This article instead, highlights the novelty of the adopted method and it should be noted that this application currently functions as a research tool. Therefore, a measuring of actual processing time has not yet been performed and hence critical processing time details are not included in this article. However, the application was run using a standard Intel 3.06 GHz processor and a 512 MB ram and hence, the processing power used was moderate.

Another important factor that is displayed first in the "Estimated Parameters Per Cell" section is the number of iterations. It shows the number of iterations the parameter estimation loop in the algorithm had to perform before producing the estimated values. In the case of Figure 1, it shows that only two iterations have occurred, when simulating the BP SX 150 solar panel. This fact actually implies that the algorithm is quite efficient in estimating the parameters quickly and more about this is discussed in the sections to follow.

Once the parameter estimation is completed, the facility to generate the solar panel output characteristic curves and tracking the maximum power point is enabled. For this, the user only has to set an irradiance and a cell temperature level from the two sliders visible in the bottom left of the interface. Then, almost instantaneously, the two figures showing the solar panel Volt - Ampere characteristic and the power profile are generated while displaying the tracked maximum power point data in the "Maximum Power Point Data" section. This interface is ideal for users to play around with fast and random variations of solar irradiance and cell temperature and study the effect on solar panel outputs since irradiance and temperature can be easily varied by sliding and the algorithm is fast enough to respond to quick changes.

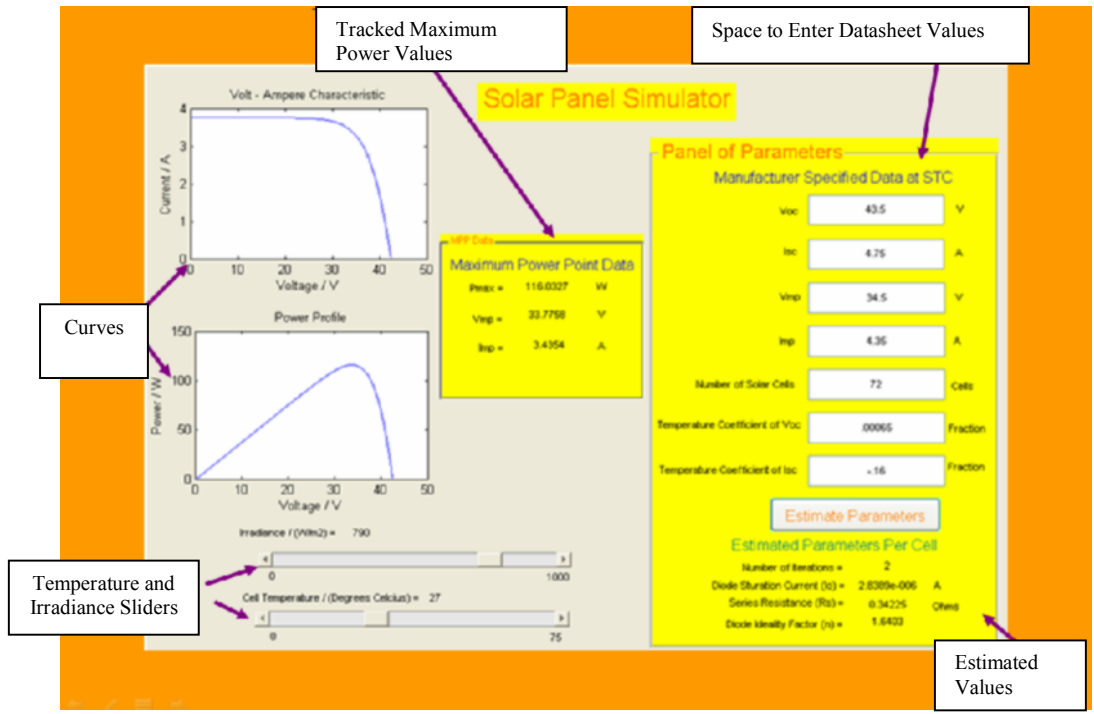

**Figure 1.** Designed software interface in MATLAB

### 4. Overview of the algorithm developed for simulating solar panels

The underlying algorithm of this application is elaborated in the flowchart shown in Figure2. As explained in section 3, the process starts with the user entering the manufacturer specified data of a solar panel, similar to those in Table 1. After that, the process is carried out based on equation (1). As mentioned in the introduction, some unknown parameters of equation (1) should be estimated before equation (1) can be used [1, 7]. These parameters are the diode ideality factor ($n$), diode reverse saturation current ($I_0$) and the series resistance ($R_s$) [1, 7].

Two major salient features of the method proposed in this article are estimating $n$ and $R_s$ in different ways to the methods proposed in literature. More about these methods are discussed in section 5. These novel methods of estimating $n$ and $R_s$ have actually enabled the development of this software application in such a way that is sufficiently fast for practical usage.

Once $n$, $I_0$ and $R_s$ are estimated, it is a matter of taking the environmental conditions in to considerations. Out of many environmental conditions that may affect the output of solar panels, the solar irradiance and cell temperature happen to be the most dominant [1, 7]. Therefore, the developed application takes only solar irradiance in W/m$^2$ and cell temperature in °C into consideration. However, it may be possible to improve this application to facilitate considering more sophisticated environmental conditions. It should be noted that the environmental conditions pose major effects on $I_0$, $I_{sc}$ and the open circuit voltage of the solar panel ($V_{oc}$) [1, 7]. Hence, the method proposed in this article models those effects by means of simple equations as discussed in section 6.

Once parameter estimation is done and environmental conditions have been taken into account, the algorithm displays the solar panel output curves and maximum power point (mpp) data to the user in a convenient manner. These results generated by the application proved to be in good accordance with the results provided by solar panel manufacturers and a validation is elaborated in section 8.

## 5. Method of parameter estimation

The novel methods of estimating the unknown parameters *n*, $I_0$ and $R_s$ in equation (1) are described in this section.

### 5.1. Novel method of estimating *n*

From the processes available in literature for estimating the unknowns, the processes of finding *n* seem to be quite tedious. For instance, in [1] and [7] *n* has been estimated by trial and error. First, a manufacturer specified *I - V* curve of a solar panel has been taken at a certain environmental condition. This curve has been used as a reference in that method. Different *I - V* curves have then been generated using equation (1) by varying the value of *n*. Next, the *n* value of the generated *I - V* curve that fits closest to the manufacturer specified *I - V* curve has been taken as the best estimate for *n*. It is evident that *n* can also be estimated by following a method of calculating the minimum sum of squares as explained in [10]. In that work too, a manufacturer specified *I - V* curve of a solar panel has been taken as a reference. Then, curves have been generated using a more complex model equation similar to equation (1) by giving assumed values to the unknown parameters including *n*. The sum of squares of the differences between the reference curve and the generated curves are then considered. Finally, the MATLAB optimization toolbox has been used to estimate the best fitting unknown parameter values by minimizing the sum of squares of the differences between the two curves.

For the estimation to be performed in either of those ways, a manufacturer specified *I - V* curve is a definite necessity. If not, *I - V* curves of solar panels should be obtained experimentally. Hence, those methods can become very tedious. To add to that, some solar panel datasheets like [4] do not provide manufacturer specified *I - V* curves. Thus, estimating *n* of such solar panels can be extremely difficult if the methods available in literature are followed.

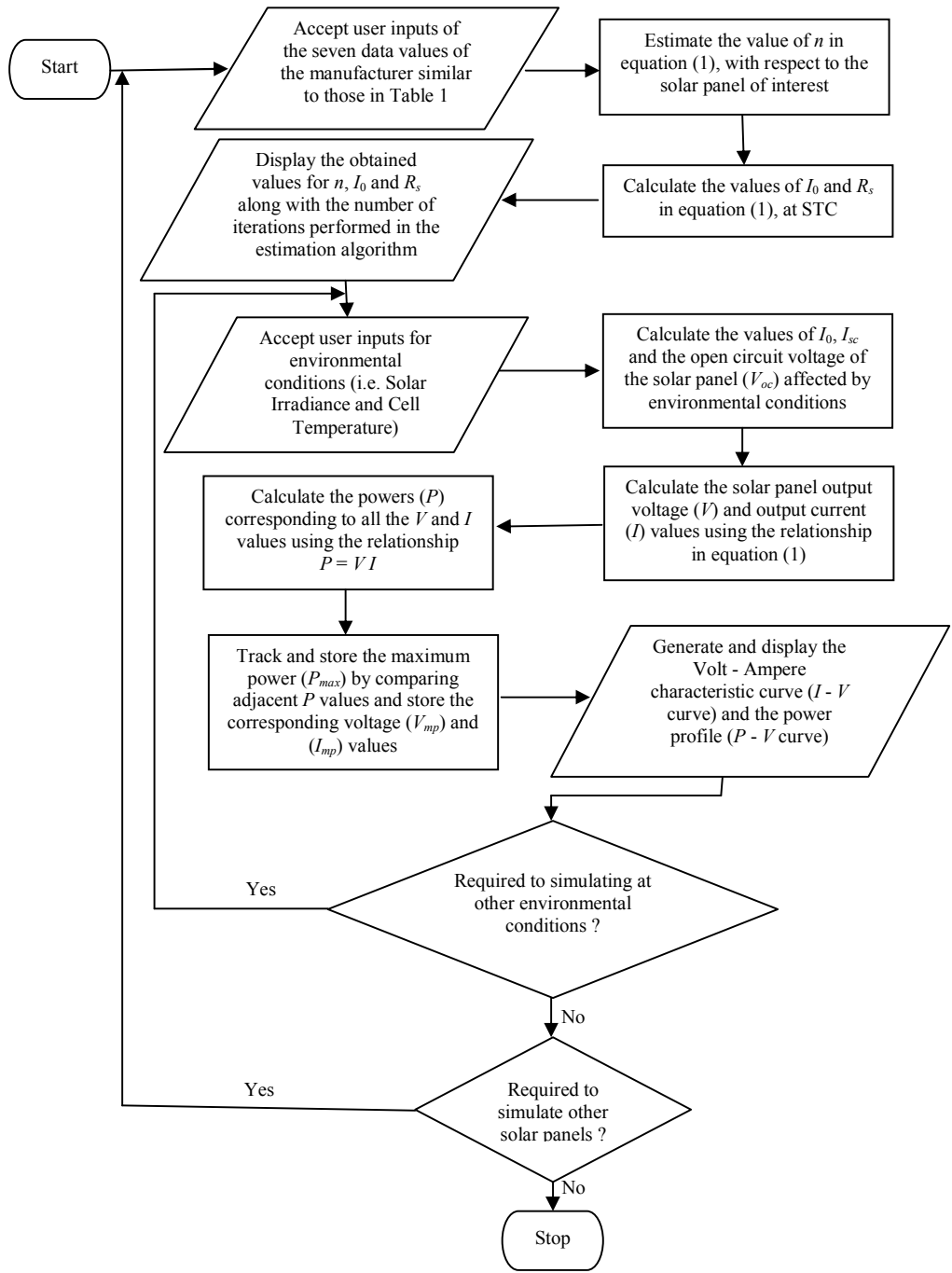

**Figure 2.** Proposed algorithm for simulating solar panels

Therefore, a simpler method of estimating *n* is proposed in this article to overcome the above mentioned difficulties. The proposed method is based on the property of *dP/dV* = 0 at the maximum power point. In the proposed method, an improvement is made by making it feasible to numerically estimate *n* by solving an analytically derived equation. Further, as mentioned in previous sections, only a few data values similar to those in Table 1 are required for the task. This indeed is a contrasting advantage of the proposed method when comparing with methods available in literature since those methods require many points of an *I - V* curve.

The actual data required for the estimation process are $V_{oc}|_{stc}$, $I_{sc}|_{stc}$, $V_{mp}|_{stc}$, $I_{mp}|_{stc}$ and *N*. Thus, *n* is estimated using the data at STC and assumed to be constant at other environmental conditions in this approach. The exact effect of environmental conditions on *n* is an ongoing research interest and therefore not addressed in this article. However, the assumption of the *n* value estimated at STC remaining constant in other conditions seems to be very reasonable as per the results obtained.

The proposed method of estimating *n* was begun by applying the open circuit condition to equation (1) at STC. At the open circuit condition, the output current *I* becomes zero and the output voltage *V* becomes equal to the open circuit voltage $V_{oc}$. But, since the open circuit condition is applied at STC, $V_{oc}$ in this case is actually equal to $V_{oc}|_{stc}$. Apart from that, it should be noted that the term *T* in equation (1) was replaced by $T|_{stc}$ which is the cell temperature at STC (i.e. 298 K) at this point. Thus, by applying the STC condition and rearranging terms, equation (2) which expresses the diode reverse saturation current at STC ($I_0|_{stc}$) was derived. It should be noted that the expression for $I_0|_{stc}$ is a function of *n* only and all the other components are constants at STC.

$$I_0\big|_{stc} = \frac{I_{sc}\big|_{stc}}{\exp\left(\dfrac{mV_{oc}\big|_{stc}}{n}\right) - 1} \qquad (2)$$

$$\text{where } m = \frac{q}{NkT\big|_{stc}}$$

Then, by applying the maximum power point conditions at STC (i.e. $V_{mp}|_{stc}$ and $I_{mp}|_{stc}$) to equation (1) and rearranging terms, a convenient equation which expresses the series resistance at STC ($R_s|_{stc}$) was obtained. That expression is shown as equation (3). Like in equation (2), it should be noted that the expression for $R_s|_{stc}$ is a function of *n* only and all the other components are constants at STC.

$$R_s\big|_{stc} = \frac{n}{mI_{mp}\big|_{stc}} \ln\left(\frac{I_{sc}\big|_{stc} - I_{mp}\big|_{stc} + I_0\big|_{stc}}{I_0\big|_{stc}}\right) - \frac{V_{mp}\big|_{stc}}{I_{mp}\big|_{stc}} \qquad (3)$$

Since this process of estimating *n* is based on the property of *dP/dV* = 0 at a maximum power point on the power profile, to proceed further equation (4) was required. This equation was derived starting from the standard relationship $P = VI$ where $I$, $V$ and $P$ are the solar panel output current, output voltage and output power respectively. By differentiating $P = VI$ with respect to $V$, the result $\frac{1}{V}\frac{dP}{dV} = \frac{dI}{dV} + \frac{I}{V}$ was obtained and by applying maximum power point conditions at STC to that result, equation (4) could be obtained since $\left(\frac{dP}{dV}\right)_{mp}\bigg|_{stc} = 0$.

$$\left(\frac{dI}{dV}\right)_{mp}\bigg|_{stc} + \frac{I_{mp}\big|_{stc}}{V_{mp}\big|_{stc}} = 0 \tag{4}$$

Next, the $dI/dV$ term in equation (4) had to be eliminated and therefore, equation (5) was derived by differentiating (1) with respect to $V$ and rearranging terms.

$$\left(\frac{dI}{dV}\right)_{mp}\bigg|_{stc} = \frac{-mI_0\big|_{stc}\exp\left\{\dfrac{m\left(V_{mp}\big|_{stc} + I_{mp}\big|_{stc}R_s\big|_{stc}\right)}{n}\right\}}{n + mI_0\big|_{stc}R_s\big|_{stc}\exp\left\{\dfrac{m\left(V_{mp}\big|_{stc} + I_{mp}\big|_{stc}R_s\big|_{stc}\right)}{n}\right\}} \tag{5}$$

Thus, elimination of $\left(\frac{dI}{dV}\right)_{mp}\bigg|_{stc}$ in equation (4), using equation (5), resulted in what is shown in equation (6).

$$\frac{nI_{mp}\big|_{stc} + \left(I_{mp}\big|_{stc}R_s\big|_{stc} - V_{mp}\big|_{stc}\right)mI_0\big|_{stc}\exp\left\{\dfrac{m\left(V_{mp}\big|_{stc} + I_{mp}\big|_{stc}R_s\big|_{stc}\right)}{n}\right\}}{\left[n + mI_0\big|_{stc}R_s\big|_{stc}\exp\left\{\dfrac{m\left(V_{mp}\big|_{stc} + I_{mp}\big|_{stc}R_s\big|_{stc}\right)}{n}\right\}\right]V_{mp}\big|_{stc}} = 0 \tag{6}$$

Since the denominator of equation (6) would be finite, it was possible to equate the numerator of equation (6) to zero as shown in equation (7).

$$nI_{mp}|_{stc} + \left(I_{mp}|_{stc} R_s|_{stc} - V_{mp}|_{stc}\right) mI_0|_{stc} \exp\left\{\frac{m\left(V_{mp}|_{stc} + I_{mp}|_{stc} R_s|_{stc}\right)}{n}\right\} = 0 \quad (7)$$

It was then intended to eliminate the exponential term in equation (7). For that, equation (8) which was derived by rearranging terms of equation (3) was used.

$$\exp\left\{\frac{m\left(V_{mp}|_{stc} + I_{mp}|_{stc} R_s|_{stc}\right)}{n}\right\} = \frac{I_{sc}|_{stc} - I_{mp}|_{stc} + I_0|_{stc}}{I_0|_{stc}} \quad (8)$$

Equations (7) and (8) were then combined and the resulted expression is given as equation (9).

$$nI_{mp}|_{stc} + m\left(I_{mp}|_{stc} R_s|_{stc} - V_{mp}|_{stc}\right)\left(I_{sc}|_{stc} - I_{mp}|_{stc} + I_0|_{stc}\right) = 0 \quad (9)$$

The next task was to eliminate $R_s|_{stc}$ from equation (9) and that was achieved by combining equations (3) and (9). The obtained result is shown in equation (10).

$$nI_{mp}|_{stc} + \left(I_{sc}|_{stc} - I_{mp}|_{stc} + I_0|_{stc}\right)\left\{n\ln\left(\frac{I_{sc}|_{stc} - I_{mp}|_{stc} + I_0|_{stc}}{I_0|_{stc}}\right) - 2mV_{mp}|_{stc}\right\} = 0 \quad (10)$$

Since $I_0|_{stc}$ is a function of only *n* as mentioned before, it was evident that the left hand side of equation (10) was a function of only *n* at STC. Therefore, the expression on the left hand side of equation (10) was denoted as *f(n)*. Thus, the novelty in the method of estimating *n* comes through finding *n* by solving *f(n)* = 0 where

$$f(n) = nI_{mp}|_{stc} + \left(I_{sc}|_{stc} - I_{mp}|_{stc} + I_0|_{stc}\right)\left\{n\ln\left(\frac{I_{sc}|_{stc} - I_{mp}|_{stc} + I_0|_{stc}}{I_0|_{stc}}\right) - 2mV_{mp}|_{stc}\right\} \quad (11)$$

This shows the simplicity of the novel method in comparison with the methods available in literature which use large amounts of data. But the proposed method requires only the STC values provided by the manufacturer in the solar panel data sheet and the value of *n* can be obtained by solving *f(n)* = 0. However, first of all it needed to be verified whether an acceptable solution exists for *f(n)* = 0. This was examined by plotting *f(n)* by sweeping *n* from zero to ten by using STC values of different solar panels whose datasheets are [2 - 4]. The results were encouraging as all the solar panels exhibited a solution between one and three. However, in [1] it is mentioned that $n$ should usually exist between one and two but this research shows contradicting results as $n$ values of some solar panels are produced to be between two and three. This phenomenon is still to be explained and provides researchers space to think about. Therefore it is not intended to provide conclusions about that matter in this article. However, the value of $n$ falling between two and three did not show any significant effect on the accuracy of the final results of the respective solar panels, generated by the application.

To elaborate the existence of a solution for $f(n) = 0$, Figures 3 and 4 were generated by plotting $f(n)$ against $n$ for the BP SX 150 solar panel. The STC data provided in Table 1 were used.

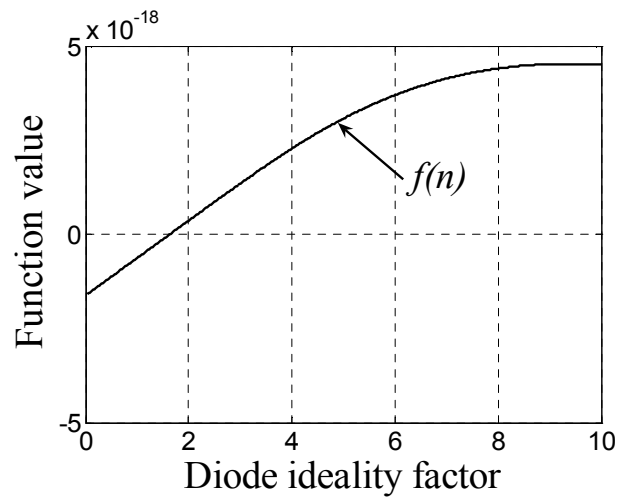

**Figure 3.** Variation of $f(n)$ against $n$ for the BP SX 105 solar panel

Figure 4 shows a zoomed view of the region where the solution for $f(n) = 0$ exists in Figure 3 and the solution in this case was found to be 1.64 which is acceptable.

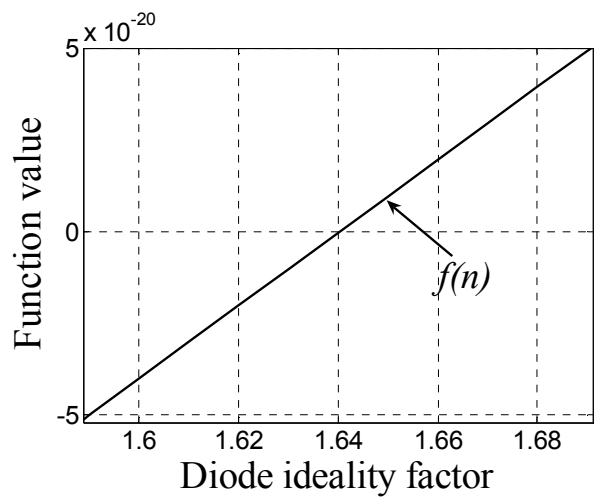

**Figure 4.** Zoomed view of the region where the solution exists for $f(n) = 0$ for the BP SX 150 solar panel

As per Figure 4, the function *f(n)* happened to be a smooth differentiable curve and therefore an effort was made in iteratively solving the equation using the Newton Raphson method as shown in equation (12).

$$n_{i+1} = n_i - \frac{f(n_i)}{f'(n_i)} \tag{12}$$

Here $n_{i+1}$ is the solution obtained iteratively after performing *i* iterations by setting some initial *n* value symbolized by $n_1$. The term *f'(n)* is the first derivative of *f(n)* with respect to *n* and the expression derived is shown in equation (13).

$$f'(n) = I_{mp}|_{stc} + \frac{dI_0|_{stc}}{dn}\left\{n \ln\left(\frac{I_{sc}|_{stc} - I_{mp}|_{stc} + I_0|_{stc}}{I_0|_{stc}}\right) - 2mV_{mp}|_{stc}\right\} + $$

$$\left(I_{sc}|_{stc} - I_{mp}|_{stc} + I_0|_{stc}\right)\left\{\ln\left(\frac{I_{sc}|_{stc} - I_{mp}|_{stc} + I_0|_{stc}}{I_0|_{stc}}\right) - \frac{n\left(I_{sc}|_{stc} - I_{mp}|_{stc}\right)}{\left(I_{sc}|_{stc} - I_{mp}|_{stc} + I_0|_{stc}\right)I_0|_{stc}}\left(\frac{dI_0|_{stc}}{dn}\right)\right\} \tag{13}$$

The term $dI_0|_{stc} / dn$ in equation (13) is the derivative of $I_0|_{stc}$ expressed in equation (2), with respect to *n*. The expression obtained for $dI_0|_{stc} / dn$ is given in equation (14).

$$\frac{dI_0|_{stc}}{dn} = \frac{mV_{oc}|_{stc} I_{sc}|_{stc} \exp\left(\frac{mV_{oc}|_{stc}}{n}\right)}{n^2\left\{\exp\left(\frac{mV_{oc}|_{stc}}{n}\right) - 1\right\}^2} \tag{14}$$

By writing a simple program to execute the iterating procedure in equation (12), it was possible to obtain the solution for *n* in just two iterations for the BP SX 150 solar panel. This was tested with other solar panels [2 - 4] as well and in all the cases a solution could be obtained in six or less iterations. This can be considered as a positive indication of the efficiency of the process of estimating *n* with respect to number of iterations performed. It should be noted that the initial value for *n* was set to be one for the estimation process as that showed to be a reasonable initial guess for *n*, which goes together with the results of this research and literature.

As mentioned in the initial paragraphs of this section, the *n* value estimated in this way is thereafter considered to be constant at all environmental conditions. This consideration proved to be reasonable enough as per the obtained results presented in the sections to follow. However, there is room for interested researchers to suggest methods to model the effect of environmental conditions on *n*, if there are any.

### 5.2. Novel method of estimating $R_s$

After estimating $n$, it was just a matter of calculating the corresponding value for $R_s$ by using equation (3). But, researchers will notice that equation (3) is expressed only for $R_s$ at STC. It should be stated that in this method, no further modeling is done to $R_s$ with respect to environmental conditions. This means, the estimated value for $R_s$ at STC ($R_s|_{stc}$) is taken as a constant irrespective of environmental conditions. Scholars might argue that this method is erroneous but it can be showed that this approach, although not flawless, does not make a significant effect on the end results.

The reason for not modeling the effect of environmental conditions on $R_s$ was simply the complexity of the process. Several publications such as [1] and [7] state that $R_s$ is heavily dependent on the gradient of I - V curves near the open circuit points. A clear description about that along with modeling equations is provided in [1] and [7] for interested scholars. However, the gradients of I - V curves are not known at the beginning and no clear evidence was found about any technique of calculating them numerically. That is an indication for the fact that those gradients are found experimentally. But, since it is intended to propose a simple PC based simulation application in this article, experimental methods of obtaining gradients of I - V curves near open circuit points was not adopted. Hence, this again creates room for interested researchers to suggest simple methods of modeling environmental effects on $R_s$ in a numerical simulation technique such as this work.

### 5.3. Method of estimating $I_0$

Similar to $R_s$, $I_0$ at STC too was estimated by substituting the estimated $n$ value to equation (2). But, unlike in the cases of $n$ and $R_s$, environmental effect on $I_0$ was taken into consideration in this approach and the method used is explained in section 6.

### 5.4. Estimated unknown values for the BP SX 150 solar panel

The unknown values estimated for the BP SX 150 solar panel through the proposed method are shown in Table 2. The processes described in sections 5.1, 5.2 and 5.3 were followed and the manufacturer specified data in Table 1 were used. An indication of the dependence of those values on environmental conditions is also included for clarity.

Table 2. Estimated unknown values for the BP SX 150 solar panel and their dependence on environmental conditions

| Symbol | Parameter | Value | Dependence on environmental conditions |
|--------|-----------|-------|----------------------------------------|
| $n$ | Diode ideality factor | 1.64 | Taken as constant for all conditions |
| $R_s$ | Series resistance | 0.342 Ω | Taken as constant for all conditions |
| $I_0$ | Diode reverse saturation current | 2.83 μA | Effect of environmental conditions are modeled |

## 6. Modeling effects of environmental conditions

Basically, the environmental conditions make significant effects on the open circuit voltage ($V_{oc}$) and short circuit current ($I_{sc}$) of solar panels [1, 7]. In this approach, the two most dominant environmental conditions that effect solar panels, which are solar irradiance and cell temperature were taken into consideration. These effects were modeled using standard formulas shown in equations (15) and (16) [7].

$$V_{oc}|_{G,T} = V_{oc}|_{stc} + b\left(T - T|_{stc}\right) + \frac{nNkT}{q}\ln(G) \tag{15}$$

$$I_{sc}|_{G,T} = I_{sc}|_{stc} G\left[1 + a\left(T - T|_{stc}\right)\right] \tag{16}$$

In equations (15) and (16), $G$, $T$, $a$ and $b$ are the desired solar irradiance in kW/m², desired cell temperature in Kelvin, the manufacturer specified fractional temperature coefficient of $I_{sc}$ and the manufacturer specified fractional temperature coefficient of $V_{oc}$ respectively. It should also be noted that $T|_{stc}$ is the temperature at STC and it too should be taken in Kelvin at this point.

As mentioned in section 5.3, the effect of environmental conditions on $I_0$ was also modeled. It has been found that $I_0$ is mainly affected by cell temperature [1, 7]. A standard equation for modeling the temperature effect is provided in [1] and [7], but to make things simpler, in this method, equation (2) itself was used for the purpose as shown in (17).

$$I_0|_{G,T} = \frac{I_{sc}|_{G,T}}{\exp\left(\frac{m_T V_{oc}|_{G,T}}{n}\right) - 1} \tag{17}$$

where $m_T = \dfrac{q}{NkT}$ and $T$ is a desired temperature in Kelvin.

This can be considered as another salient features of this approach since the method of modeling the temperature effect on $I_0$ is different from the traditional method shown in [1, 7] and conceptually, using equation (17) is correct since it contains modeled $V_{oc}$, modeled $I_{sc}$ and the actual temperature in $m_T$.

Thus, after finding values $V_{oc}|_{G,T}$, $I_{sc}|_{G,T}$ and $I_0|_{G,T}$ which are modeled by taking environmental conditions into consideration, the output Volt - Ampere curves (I - V curves) and the power profiles (P - V curves) of commercial solar panels, similar to those visible on the software interface shown in Figure 1, were generated using equation (1) as explained in section 7.

## 7. Generating output curves of solar panels

Since two different curves (i.e. I - V curve and the P - V curve) were generated, the methods of generation of the two curves are explained separately.

### 7.1. Generating Volt - Ampere curves (I - V curves)

Once the procedures described in sections 5 and 6 were carried out., an array of values for the output current ($I$) was defined from zero to the required $I_{sc}|_{G,T}$ value. Then, a code was written to calculate a corresponding array containing values of the output voltage ($V$) using equation (18) which is only a rearrangement of terms in equation (1).

$$V = \frac{n}{m_T} \ln\left(\frac{I_{sc}|_{G,T} - I + I_0|_{G,T}}{I_0|_{G,T}}\right) - IR_S \tag{18}$$

Finally, the I - V curves were generated by plotting the $I$ arrays against $V$ arrays.

### 7.2. Generating the power profiles (P - V curves)

Power profiles were simply generated using the relationship $P = VI$ by calculating new $P$ arrays by multiplying corresponding elements in $V$ and $I$ arrays and plotting them against $V$ arrays.

## 8. Results and validation

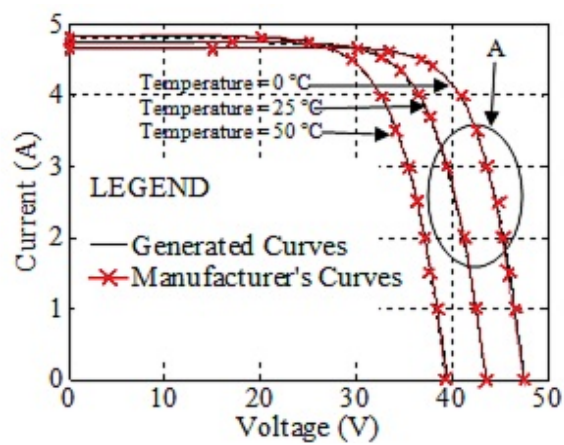

Figure 5. Comparison between the generated I - V curves and the manufacturer specified curves of the BP SX 150 solar panel at different cell temperatures at 1000 W/m² solar irradiance

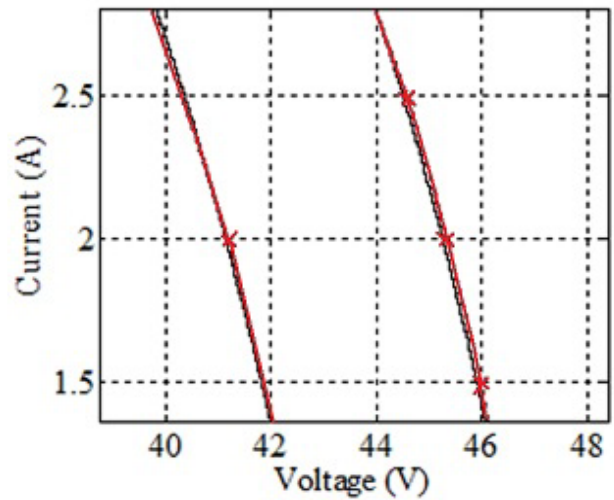

**Figure 6.** Zoomed view of region 'A' marked on Figure 5

The next task was to validate the results generated by the developed application. For that, a comparison between the commercial solar panel I - V curves generated by the application and the I - V curves specified by the respective solar panel manufacturers was performed. The BP SX 150 and BP 3210 N solar panels were selected for the purpose. The manufacturer of the BP SX 150 solar panel had provided I - V curves at different cell temperatures at 1000 W/m$^2$ solar irradiance. On the other hand, the BP 3210 N solar panel manufacturer had provided I - V curves at different irradiance levels at a cell temperature of 25 °C. Therefore, these two solar panels were ideal to validate the simulation results with respect to both cell temperature and solar irradiance. The manufacturer specified I -V curves were obtained from the respective solar panel datasheets [2, 3]. The obtained results are shown in Figures 5, 6 and 8. It should be noted that a legend is not marked on Figures 6 and 8 to maintain clarity of the figures and therefore they use the same legend in Figure 5.

The difference between the generated curves and manufacturer specified curves was found to be very small for the BP SX 150 solar panel as per Figure 6. This aspect was tested with not only the BP SX 150 solar panel, but also with many other panels and the results proved to be equally good. This can be considered as a validation of the accuracy of the generated results through the proposed software application. The generated power profiles of the BP SX 150 solar panel corresponding to the simulated curves in Figure 5 are shown in Figure 7. The profile corresponding to 25 °C happens to be the power profile at STC for the BP SX 150 solar panel as the considered irradiance in Figure 5 is 1000 W/m$^2$. Table 1 specifies the STC voltage at maximum power ($V_{mp}|_{stc}$) as 34.5 V. The BP SX 150 solar panel data sheet [2] gives the approximate maximum power at STC as 150 W. Thus, Figure 7 shows the exact maximum power point and this too highlights the correctness of the results produced by the proposed application. However, a point wise comparison on the power profiles could not be performed unlike with the I - V curves since manufacturer's data could not be obtained. But, the important factor was obtaining the maximum power point at STC correctly since practically solar panels are operated at maximum power point conditions [17-19].

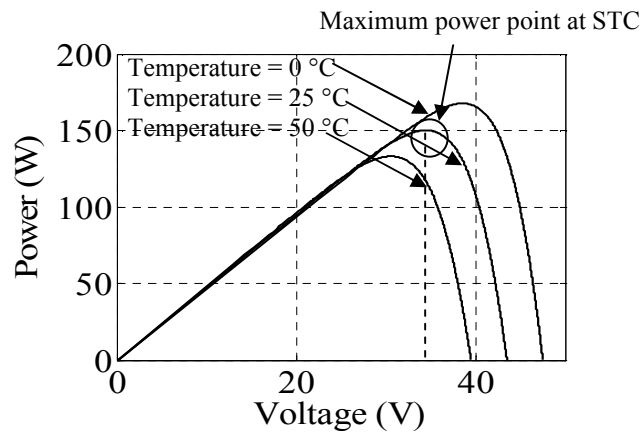

**Figure 7.** Power profiles generated for the BP SX 150 solar panel through the software application at different cell temperatures at 1000 W/m$^2$ solar irradiance

Similar to Figure 5, Figure 8 exhibits a comparison between I - V curves generated by the software application and manufacturer's curves of the BP 3210 N solar panel. The difference in this case from the previous one is that the curves being generated at different solar irradiance levels by keeping the cell temperature at 25 $^o$ C. Therefore, Figure 8 can be considered as a validation of the results generated by the software application at different irradiance levels. The manufacturer specified I - V curves for this purpose were obtained from the BP 3210 N solar panel data sheet [3]. Once again an appreciable accordance between the simulated results and manufacturer specified results were observed.

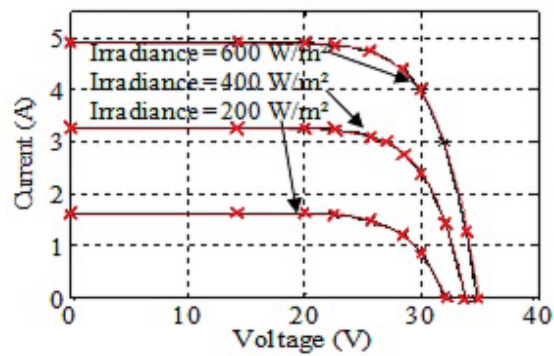

**Figure 8.** Comparison between the generated I - V curves and the manufacturer specified curves of the BP 3210 N solar panel at different irradiance levels at 25 $^o$ C cell temperature

## 9. Conclusions

According to the results elaborated in section 8, the proposed software application proves to produce commendably accurate results. The application is based on a method of estimating unknowns of the standard single diode solar cell model, corresponding to commercial solar panels. The proposed method is associated with three main salient features. The first feature is estimating the diode ideality factor by solving an analytically derived equation in the form of $f(n) = 0$. A solution with a tolerance of ± 0.0001 for $f(n) = 0$ could be produced through six or less iterations for all the tested commercial solar panels. The usage of Newton Raphson method for the purpose could have played a part in reducing the number of iterations. This is an indication of the crude efficiency of the proposed algorithm to estimate $n$ in terms of the number of iterations. The second salient feature is calculating the $R_s$ value using equation (3). This approach is clearly different from the ones available in literature [1, 7] and it proved to have a positive effect towards compromising simplicity and accuracy of this application. The third salient feature of this method is modeling the effect of temperature on $I_0$ directly using equation (2). This too played a great role in reducing the complexity of this method.

A user friendly software interface has been developed to facilitate easy interaction and comfortable visualization of solar panel I - V and P - V curves. The major ease provided to the user is the minimal input data requirement. The only data required to run the whole process are the seven data similar to those in Table 1 and those data can be found very easily by referring to any solar panel data sheet. The environmental conditions in which the output curves should be generated can be easily given by the user through the interface. The application is capable of estimating unknowns, generating output curves and tracking the maximum power point in just a matter of a few fractions of a second, when executed using moderate processing power. This is an indication of the potential of this method to be used in real time applications.

However, no measurement on precise processing time was taken or an effort was made to optimize the algorithm in terms of software engineering. This application therefore is by no means optimized up to commercial standards and is currently at the stage of a flexible research tool. This creates space for researchers to make an effort on increasing the efficiency and optimizing this application and its underlying algorithm. This article could be the ideal reference for that since all derivations of equations used are provided along with a detailed description of the methodology followed. With reasonable optimization, the proposed method could lead to produce real time, efficient, commercial solar emulators and enhance the field of innovative software development in relation to photovoltaic technology. Apart from that, it may be possible to improve the proposed application by introducing methods to model the losses associated with hardware solar panel setups by adopting a modeling technique similar to that in [20].

A few limitations exist currently in this method since modeling the effect of environmental conditions on $n$ and $R_s$ is not possible. This should undoubtedly be affecting the accuracy of the final results to at least a minor extent. Therefore, this process can be improved further by introducing innovative methods for modeling environmental conditions on $n$ and $R_s$. However, as mentioned in section 5, the effect of environmental conditions on $n$ is a current research interest and therefore not clear yet. But, this research highlights the requirement of verifying the actual environmental dependence of $n$ while creating space for interested researchers to suggest innovative methods to model the environmental dependence of $n$ and $R_s$. Apart from that, conditions such as moisture, wind and vibration too could be affecting the solar panel outputs. Modeling these effects will be very useful in simulating solar panels fixed on mobile frames like vehicles. But, such sophisticated simulating is not possible with the proposed application at the moment. Thus, room exists for researchers to improve the proposed method by introducing mechanisms to model miscellaneous effects such as wind and vibration on solar panels fixed on a vehicle for instance, which might result in a useful study similar to [21].

### References

[1] Akihiro Oi, "Design and simulation of photovoltaic water pumping system," A thesis presented to the Faculty of California Polytechnic State University, San Luis Obispo, pp. 5-36, September, 2005.

[2] BP Solar, BP SX 150 – 150 W Multi-crystalline Photovoltaic Module, Datasheet, http://www.abcsolar.com/pdf/bpsx150.pdf, 01/05/2012.

[3] BP Solar, 210 W Photovoltaic Module of Poly 3 – Series, Datasheet, http://www.bpsunoasis.com/images/BP3210N_4058E-2_0607.pdf, 01/05/2012.

[4] Solar panels, 5th Generation a – Si solar panels, Datasheet, http://www.freeenergyeurope.com/pdf/FEE-12-lineC.pdf, 01/05/2012.

[5] L.A. Dobrzański, L. Wosińska, B. Dołżańska, A. Drygała, "Comparison of Electrical Characteristics of Silicon Solar Cells," Journal of Achievements in Materials and Manufacturing Engineering, vol. 18, Issue 2, pp. 215 – 218, September – October, 2006.

[6] Govinda R. Timilsina, Lado Kurdgelashvili, Patrick A. Narbel, "A Review of Solar Energy Markets, Economics and Policies", Policy Research Working Paper (WPS5845) published by the World Bank Development Research Group, Environment and Energy Team, October, 2011.

[7] Walker, Geoff R. "Evaluating MPPT Converter Topologies Using a MATLAB PV Model", Australasian Universities Power Engineering Conference, AUPEC '00, Brisbane, 2000.

[8] Liliana Cortez, J. Italo Cortez, Alejandro Adorno, Ernest Cortez, Mariano Larios, "Simulation of a Photovoltaic Solar Module for the Study of the Effects of Random Changes of Solar Radiation", 12th WSEAS International Conference, Automatic Control, Modeling & Simulation, May, 2010.

[9] R.K. Nema, Savita Nema, and Gayatri Agnihotri, "Computer Simulation Based Study of Photovoltaic Cells/Modules and their Experimental Verification", International Journal of Recent Trends in Engineering, vol. 1, Issue 3, pp. 151 – 156, May, 2009.

[10] MATLAB demos, "Solar Cell Parameter Extraction From Data", http://www.mathworks.com/products/demos/shipping/elec/elec_solar_opt_m.html?product=SN, 01/05/2012.

[11] Krisztina Leban, Ewen Ritchie, "Selecting the Accurate Solar Panel Simulation Model", NORPIE/2008, Nordic Workshop on Power and Industrial Electronics, June 9-11, 2008.

[12] Mohamed Azab, "Improved Circuit Model of Photovoltaic Array", International Journal of Electrical Power and Energy Systems Engineering, vol. 2, Issue 3, pp. 185 - 188, 2009.

[13] F. Adamo, F. Attivissimo, A. Di Nisio, A. M. L. Lanzolla, M. Spadavecchia, "Parameters Estimation for a Model of Photovoltaic Panels", XIX IMEKO World Congress, Fundamentals and Applied Metrology, Lisbon, Portugal, September 6 – 11, 2009.

[14] Guillermo Martín-Segura, Joaquim López-Mestre, Miquel Teixidó-Casas, Antoni Sudrià-Andreu, "Development of a Photovoltaic Array Emulator System based on a Full-Bridge Structure", 9th International Conference, Electric Power Quality and Utilization, Barcelona, 9 – 11 October, 2007.

[15] H. Votzi, F. A. Himmelstoss, H. Ertl, "Basic Linear - Mode Solar - Cell Simulators", 35th Annual Conference, Industrial Electronics, IEEE, 2009.

[16] KLP – S Series Solar Emulator, Product Preview, http://www.kepcopower.com/klp-s.htm, 01/05/2012.

[17] Aurobinda Panda, M.K. Pathak, S.P. Srivastava, "Fuzzy Intelligent Controller for the Maximum Power Point Tracking of a Photovoltaic Module at Varying Atmospheric Conditions", Journal of Energy Technologies and Policy, vol. 1, Issue 2, pp. 18 - 27, 2011.

[18] C. Liu, B. Wu and R. Cheung, "Advanced Algorithm for MPPT Control of Photovoltaic Systems", Canadian Solar Buildings Conference, Montreal, August 20-24, 2004.

[19]     Texas Instruments, "Introduction to Photovoltaic Systems Maximum Power Point Tracking", Application report, SLVA446 – November, 2006, http:// focus.ti.com/lit/an/slva446/slva446.pdf, 01/05/2012.

[20]     Marinko Barukcic, Srete Nikolovski, Zeljko Hederi, "Estimation of Power Losses on Radial Feeder Using Minimum Electrical Measurements and Differential Evolution Method", International Journal of Soft Computing and Software Engineering (JSCSE), Vol. 2, No. 4, pp. 1 - 13, April, 2012.

[21]     Zulkifli Mohd Nopiah, Ahmad Kadri Junoh, Wan Zuki Azman Wan Muhamad, Mohd Jailani Mohd Nor, Ahmad Kamal Ariffin Mohd. Ihsan, Mohammad Hosseini Fouladi, "Linear Programming : Optimization of Noise and Vibration Model in Passenger Car Cabin", International Journal of Soft Computing and Software Engineering (JSCSE), Vol. 2, No. 1, pp. 1 - 13, January, 2012.